\documentstyle[aps,multicol,epsfig,prl]{revtex}

\begin{document}
 \input epsf
\draft \preprint{HEP/123-qed}

\title{Calculation of the incremental stress-strain relation of a polygonal packing}
\author{F. Alonso-Marroquin and H. J. Herrmann}
\address{ICA1, University of Stuttgart, \\
Pfaffenwaldring 27 \\
70569 Stuttgart, Germany}
\date{\today}
\maketitle


\begin{abstract}

The constitutive relation of the quasi-static deformation on two dimensional packed samples of polygons 
is calculated using molecular dynamic simulations. The stress values at which the system remains
stable are bounded by a failure surface, that shows a power law dependence on the pressure.  
Below the failure surface, non linear elasticity and plastic deformation are obtained, 
which are evaluated in the framework of the incremental linear theory. The results shows 
that the stiffness tensor can be directly related to the micro-contact rearrangements.
The plasticity obeys a non-associated flow rule, with a plastic limit surface 
that does not agree with the failure surface. 

\end{abstract}
\begin{multicols}{2}
\narrowtext


\section{Introduction}
\label{Intro}

The non linear and irreversible behavior of soils has been described by different
constitutive theories \cite{vermeer,kolymbas}. Here the stress - strain relation is postulated 
using a certain number of material parameters that are measured in experimental
tests.   In practice, the constitutive relations can be constructed directly,
taking samples with the same macroscopical state, and measuring in each one the 
incremental strain that results from the application of a specific stress
increment\cite{darve}. However, such tests are difficult to perform, because they require the 
fabrication of many samples with identical material properties.

Numerical simulations result as an alternative to the solution of this problem.
They allow to develop different tests on identically generated samples and provide
detailed information about micro-mechanical rearrangements during the loading process.
Usually, disks or spheres are used to capture the granularity of the sample \cite{cundall,bardet}.
Although the simplicity of their geometry allows to reduce the computer time of 
calculations, they do not provide a detailed description of realistic granular 
textures.

We present here a two dimensional discrete model that takes into account the diversity 
of shapes of the grains in the soils.  The granular samples consist of randomly generated 
polygons.  As presented by Tillemans \cite{tillemans}, The interaction between the polygons 
could be handled letting the polygons interpenetrate each other and calculating the force 
as a function of their overlap. This approach has been successfully applied to 
model different processes, like fragmentation \cite{Kun} and strain localization
\cite{tillemans}. A suitable contact force law is introduced in Sect.\ref{cf}, that attempts
to combine the  Hertz contact law with the Coulomb friction criterion for quasi-static 
deformation. 

The incremental stress-strain relation was calculated performing different stress 
increments on the same sample, and measuring the corresponding strain response. 
The stress is applied on the boundary through a flexible membrane that 
surrounds the sample.  The modeling of such a membrane, whose details are given in 
Sect. \ref{bf}, results more complex than rigid walls, 
However, it results more advantageous than walls, 
because it allows to implement a stress-controlled condition without any restriction 
in the deformation of the boundary.

\section{Model}
\label{model}

The polygons of this model are generated using a simple version of the Voronoi 
tessellation:  First we set a random point in each cell of a regular square 
lattice. Then each polygon is  constructed assigning to each point that part of 
the plane that is nearer to it than to any other  point. Each polygon is subjected
to interparticle contact forces and boundary forces that are inserted in  Newton's
equation of motion.
 
\subsection{Contact force}
\label{cf}

\begin{figure}
  \begin{center}
    \epsfig{file=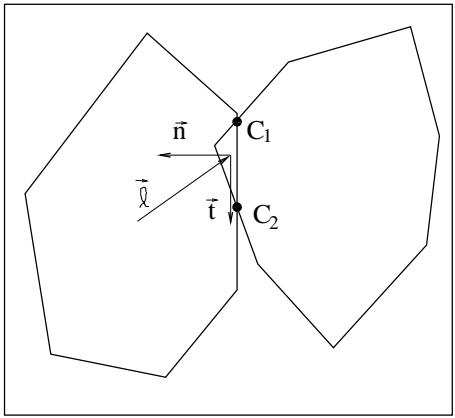,width=6.0cm,angle=0,clip=1}
   \end{center}
   \caption{ Contact surface as defined from the geometry of 
overlap.}
    \label{overlap}
\end{figure}

Usually, the interaction between two solid bodies in contact is described by a force applied 
on the flattened contact surface between them. Given two polygons in contact, such
surface is obtained from the geometrical construction shown in Fig. \ref{overlap}.
The points $C_1$ and $C_2$ result from the intersection between the edges of the polygons.
The contact surface is taken as the segment that lies between those points. 
The vector $\vec{S}=\overrightarrow{c_1 c_2}$ defines an intrinsic coordinate
system at the contact  $(\hat{t},\hat{n})$, where $\hat{t}=\vec{S}/|\vec{S}|$  and 
$\hat{n}$ is perpendicular to it. The deformation length is given by
$\delta = a/|\vec{S}|$ where $a$ is the overlap area between the polygons.
$\vec{\ell}$ is the branch vector, that connects the center of mass of the polygon to 
the point of application of the contact force, that is supposed to be the center of mass 
of the overlap area. 

The normal elastic force is taken proportional to the deformation length as 
$f^e_n=k_n\delta$; the tangential force is calculated from the simplified Coulomb friction 
law, with a single friction coefficient $\mu_{s}=\mu_{d}=\mu$. Here $\mu_s$ is the 
static and  $\mu_d$ the dynamic friction coefficient.  This tangential  force is 
implemented by an elastic spring $f^e_{t}=-k_t\xi$, where $\xi$  grows linearly with 
the tangential displacement of the contact, whenever  $|f^e_{t}|<\mu f^e_{n}$. We
used the straightforward calculation of $\xi$  proposed by Brendel \cite{Brendel}:

\begin{eqnarray}
\xi (t)=\int_{0}^{t}v_n(t')\Theta(f^e_n(t')-\mu|f^e_t(t')|)dt'
\label{eq2}
\end{eqnarray}  

\noindent
where $\Theta$ is the Heaviside function and $\vec{v}$ is the relative velocity at the
contact, that depends on the linear velocity $\vec{v}_i$ and angular velocity $\vec{\omega}_i$
of the particles in contact according to: 

\begin{equation}
\vec{v}=\vec{v}_{i}-\vec{v}_{j}-\vec{\omega}_{i}\times\vec{\ell}_{i}
+\vec{\omega}_{j}\times\vec{\ell}_{j}
\label{vr}
\end{equation}

\subsection{Boundary Forces}
\label{bf}

Let us now discuss how to apply the stress on the sample. One way to do that,
would be to apply a perpendicular force on each edge of the polygons belonging to the external contour
of the sample. Actually this does not work, because this force will act on all the fjords 
of the boundary. It produces an uncontrollable growth of cracks that with time ends up destroying
the sample. Thus, results necessary to introduce a flexible membrane in order to restrict the 
boundary points that are subjected to the external stress.

\begin{figure}
 \begin{center}
 \epsfig{file=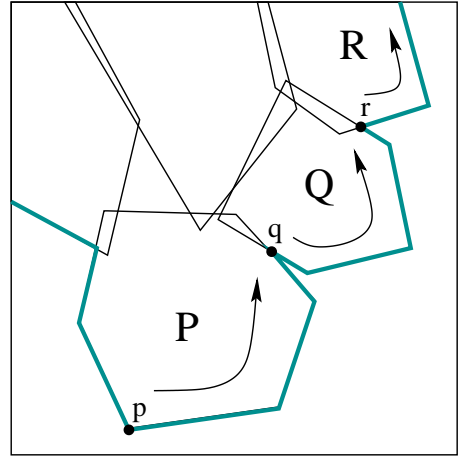,width=6.0cm,angle=0,clip=1}
 \end{center}
 \caption{Algorithm used to find the boundary.}
 \label{contour}
\end{figure}

The algorithm to identify the boundary is rather simple. The lowest vertex $p$ from all the polygons 
of the sample is chosen as the first point of the boundary list $b_1$. In Fig. \ref{contour} 
$P$ is the polygon that contains $p$, and $q\in P \cap Q $ is the first intersection 
point between the polygons $P$ and $Q$ in counterclockwise orientation with respect to $p$. 
Starting from  $p$, the vertices of $P$ in counterclockwise orientation are included in 
the boundary list until $q$ is reached. Next, $q$ is included in the boundary list. Then, the 
vertices of $Q$ between $q$ and the next intersection point $r \in Q \cap R$ in the 
counterclockwise orientation are included into the list. The same procedure is applied 
until one reaches the lowest vertex $p$ again. This is a very fast algorithm, 
because it only makes use of the contact points between the polygons, which are previously 
calculated to obtain the contact force. 

\begin{figure}
 \begin{minipage}[s]{4.0cm}
 \begin{center}
 \epsfig{file=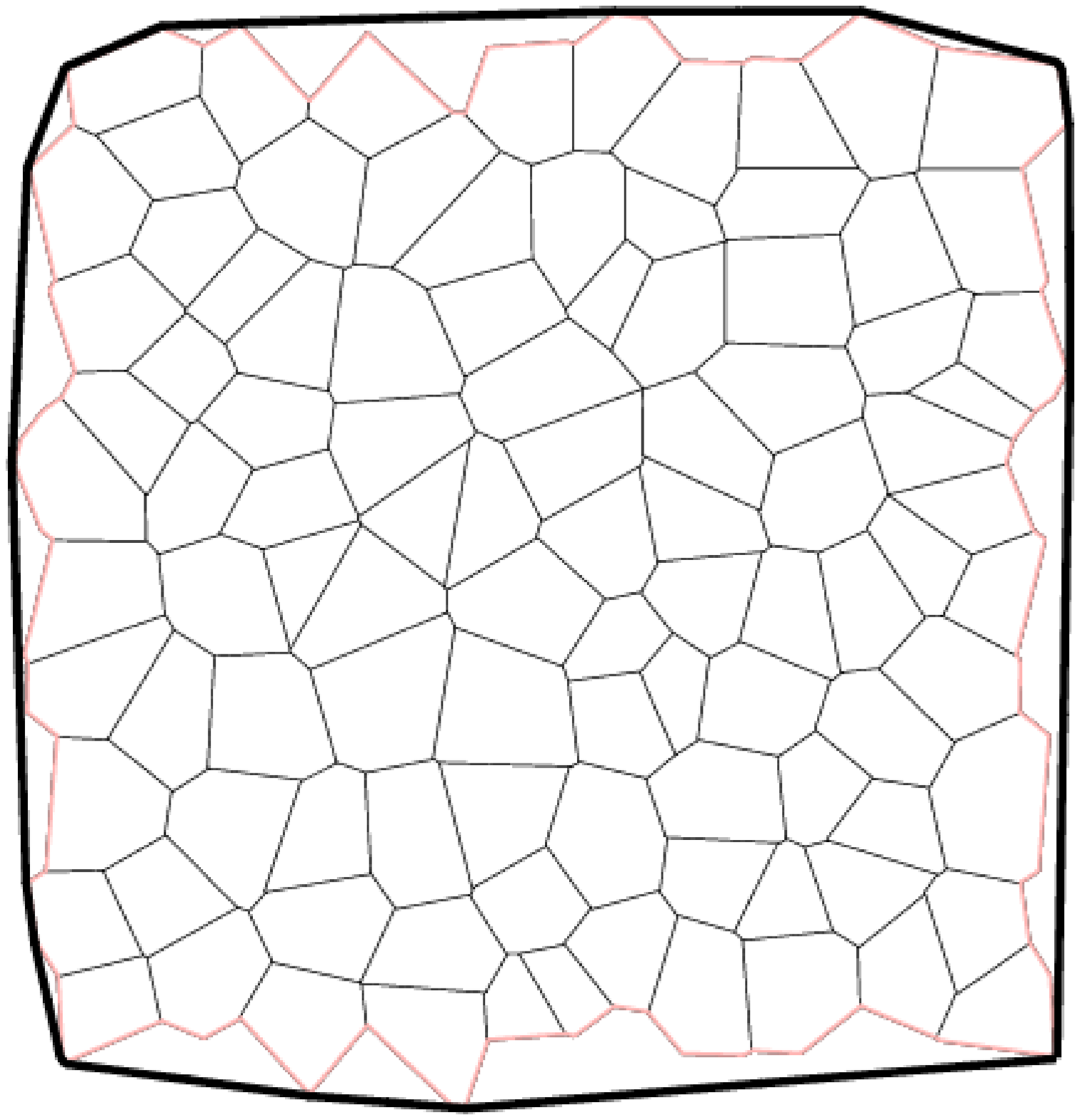,width=4.0cm,angle=0,clip=1}
 \end{center}
 \end{minipage}
 \begin{minipage}[s]{4.0cm}
 \begin{center}
 \epsfig{file=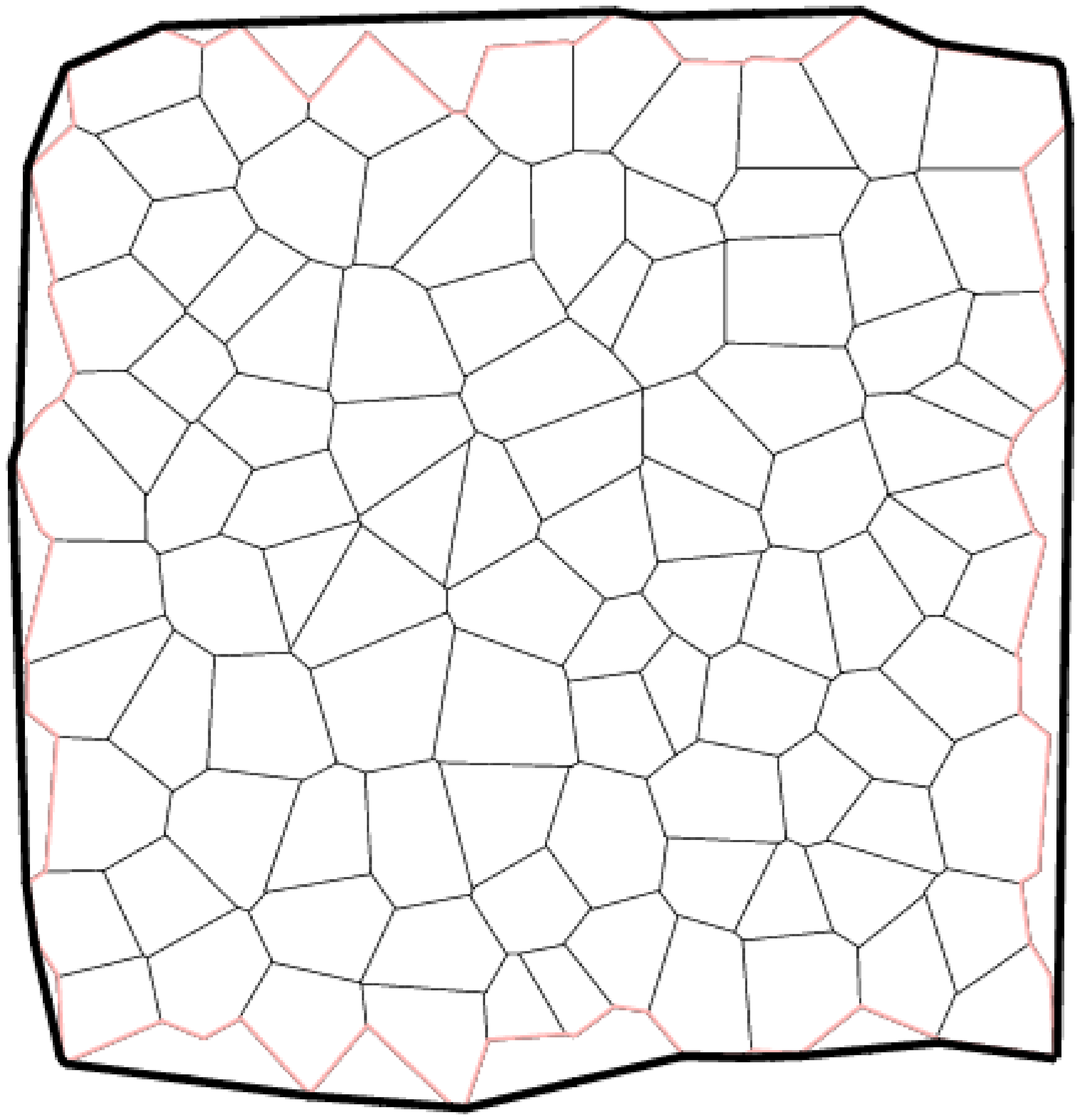,width=4.0cm,angle=0,clip=1}
 \end{center}
 \end{minipage}
\begin{minipage}[s]{4.5cm}
 \begin{center}
 \epsfig{file=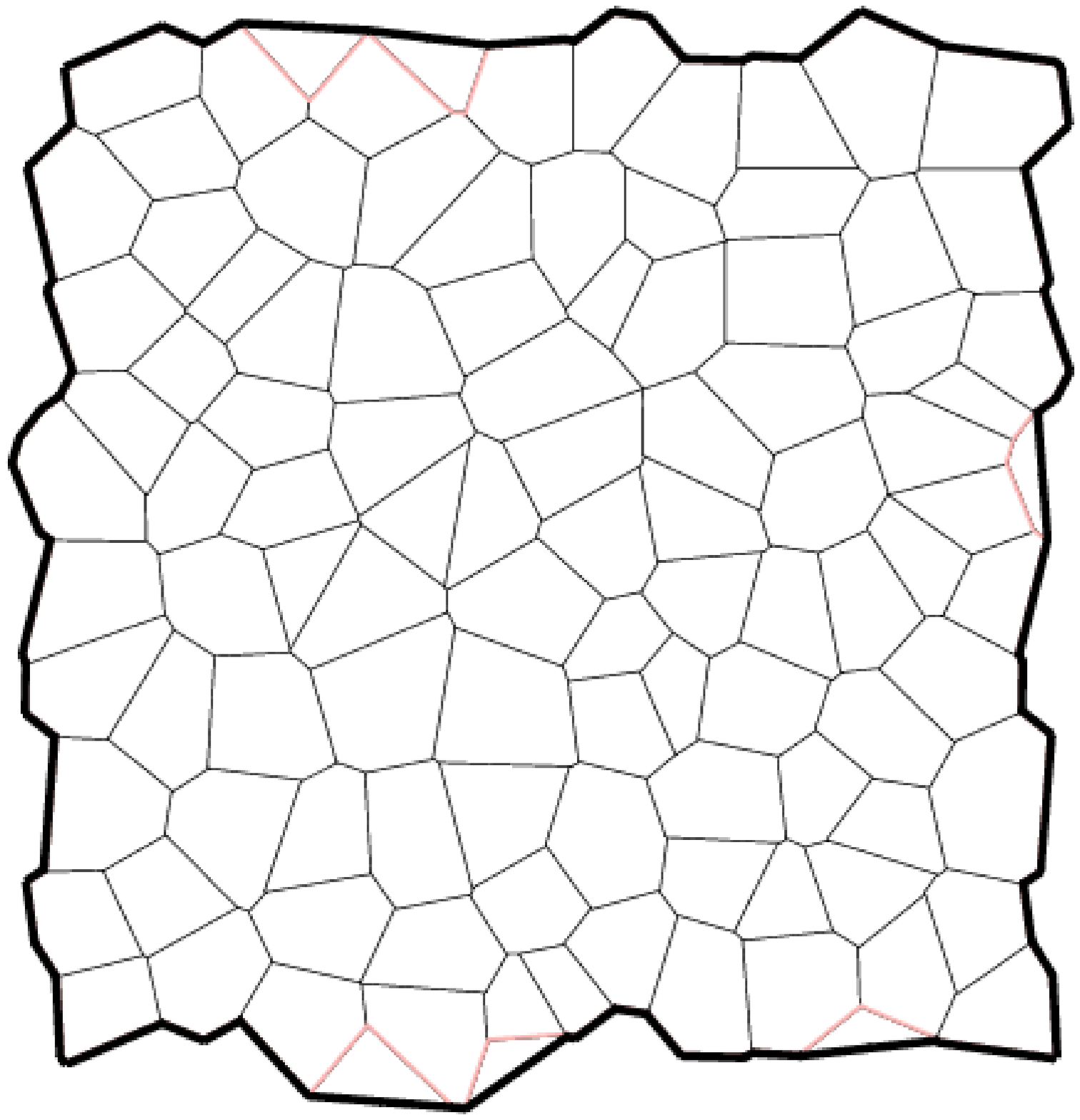,width=4.0cm,angle=0,clip=1}
 \end{center}
 \end{minipage}
 \begin{minipage}[s]{4.5cm}
 \begin{center}
 \epsfig{file=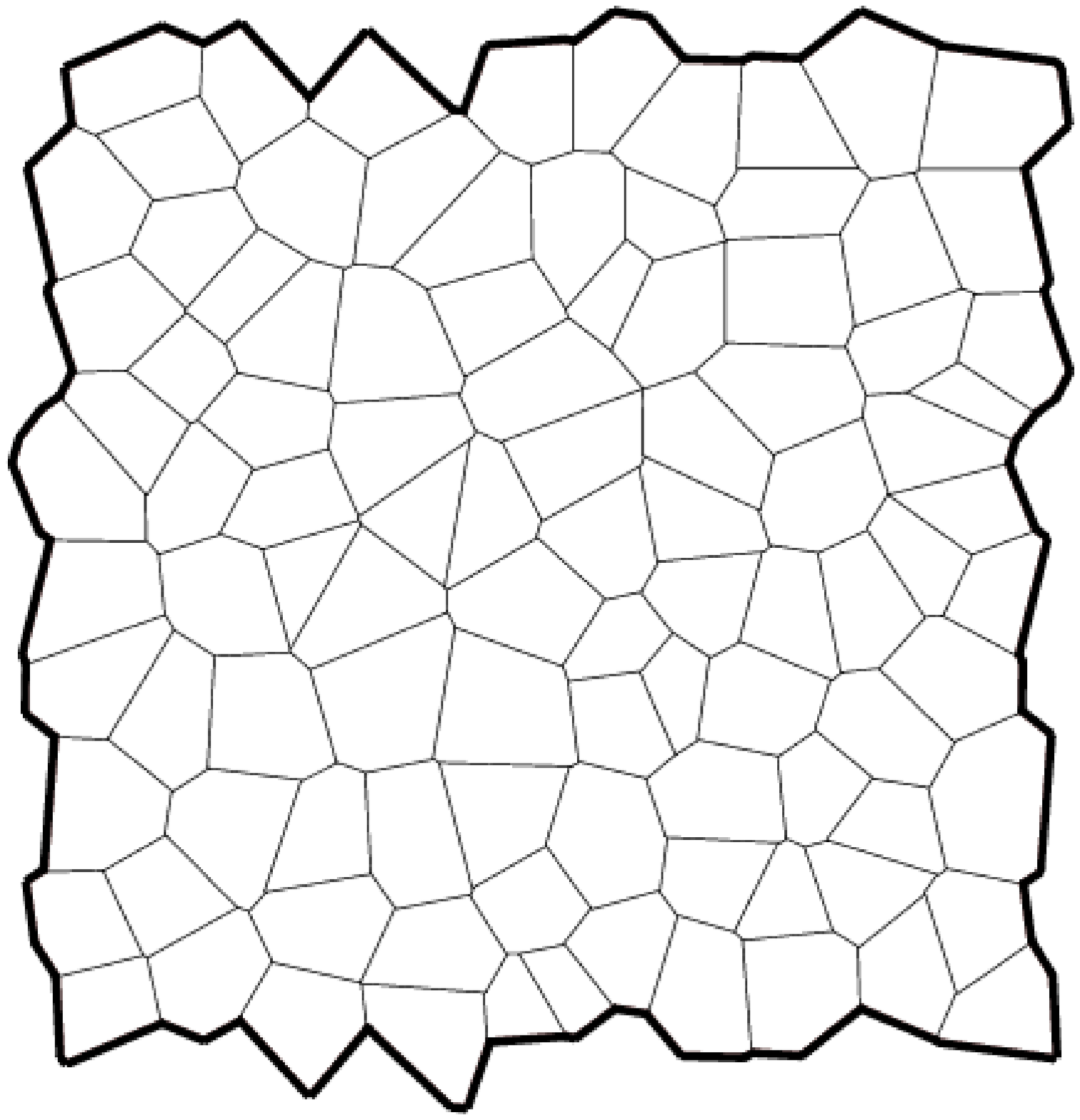,width=4.0cm,angle=0,clip=1}
 \end{center}
 \end{minipage}
\caption{Membrane obtained with threshold bending angle $\theta_{th}=\pi$, $3\pi/4$, $\pi/2$
and $\pi/4$, the first one corresponds to the minimum convex polygon that encloses the sample.}
\label{memb}
\end{figure}

The set of points that are in contact with the membrane are selected using a recursive algorithm.
It is initialized with the vertices of the smallest convex polygon that encloses the boundary.
(see Fig. \ref{memb}). The lowest point of the boundary is selected as the first vertex of 
the polygon $m_1=b_1$.  The second  one $m_2$ is the  boundary point  $b_i$ that minimizes the angle  
$\angle(\overrightarrow{b_1 b_i})$ with respect to the horizontal.   
The third one $m_3$ is the boundary point $b_i$ such that the angle  
$\angle(\overrightarrow{m_2 b_i},\overrightarrow{m_1 m_2})$ is minimal.  
The algorithm is recursively applied until the lowest vertex $m_1$ is reached again. 

The points of the boundary are iteratively included in the list ${m_i}$ using the 
bending criterion proposed by \AA str\o m \cite{astrom}: For each pair of consecutive 
vertices  of the membrane $m_i=b_i$ and $m_{i+1}=b_{j}$ we choose that point from 
the subset $\{ b_k\}_ {i \le k \le j} $ that maximizes the bending angle 
$\theta_b=\angle(\overrightarrow{b_k b_i},\overrightarrow{b_k b_{j}})$. 
This point is included into the list, whenever $\theta_b\ge\theta_{th}$. Here 
$\theta_{th}$ is a threshold angle for bending. This algorithm is  repeatedly applied until 
there are not more points  satisfying such bending condition.  

The final result gives a set of segments  $\{\overrightarrow{m_i m}_{i+1}\}$
lying on the boundary of the sample. In order to apply the boundary forces, 
those segments are divided into two groups: A-type segments are those that 
coincide with an edge of a boundary polygon;  B-type segments connect the vertices 
of two different boundary polygons. 

0n each segment of the membrane $\overrightarrow{m_i m}_{i+1}$ a force $f_i = \sigma_i N_i$ 
is applied, where $\sigma_i$ is the local stress and  $N_i$ is the $90^0$ counterclockwise
rotation of $\overrightarrow{m_i m}_{i+1}$. 
This force is transmitted to the polygons in contact with it: if the segment is
A-type, this force is applied in its midpoint; If the segment is B-type, half of the
force is applied at each one of the vertices connected by this segment. 

\subsection{Molecular dynamic simulation}

Before we implement the numerical solution of Newton's equations it is convenient to make 
a dimensional analysis of the parameters. In such way we can keep the scale invariance of the 
model and reduce the parameter to a minimum of adimensional constants. All the polygons are
supposed to have the same density. The mass $m_i$ of each polygon is measured in units of the mean 
mass $m_0$ of the Voronoi tessellation. The time is measured in fractions of the total
loading time $t_0$. The evolution of the position $\vec{x}_i$ and the orientation 
$\varphi_i$ of the $i-th$ polygon is governed by the equations of motion:

\begin{eqnarray}
\lambda^2 m_i\ddot{\vec{x}}_i+\sum_{c}\vec{f^c_i}
+\sum_{c_{b}}\frac{\sigma^b_i}{k_n}\vec{f}^b_i  &=&0 \nonumber\\
\lambda^2 I_i\ddot{\varphi}_{i}+\sum_{c}\vec{\ell}^c_i\times\vec{f^c_i}
+\sum_{c_{b}}\frac{\sigma^b_i}{k_n}\vec{\ell}^b_i\times\vec{f}^b_i&=&0 
\label{dm}
\end{eqnarray}

The sums go over all those particles and boundary segments that are in contact with the $i-th$
polygon. The interparticle contact forces $\vec{f^c_i}$ and boundary forces $\vec{f^b_i}$ are 
given by:

\begin{eqnarray}
\vec{f^c_i}&=&(\delta^c_i+\lambda\gamma m v^c_n)\hat{n}^c_i 
+\zeta(\xi^c_i - \lambda\gamma m v^c_t)\hat{t}^c_i \nonumber \\
\vec{f^b_i}&=&N^b_i -\lambda \gamma m_i \vec{v_i}
\label{dm2} 
\end{eqnarray}
 
Here $\delta^c_i$ and $\xi^c_i$ are the deformation length and the tangential displacement
of the contact, which were defined in Sect. \ref{cf}; $\sigma^b_i$ is the stress applied
on the boundary segment $T^b_i$, defined in Sect. \ref{bf}. Artificial viscous terms
must be included in Eq. (\ref{dm2}) to keep the stability of the numerical solution 
and reduce the acoustic waves generated during the loading  process. $\vec{v}^c$ is the relative
velocity at the contact (Eq. (\ref{vr})) and $m = (1/m_i + 1/m_j)^{-1}$ is the effective mass of 
the two polygons in contact.

There are four microscopic parameters in the model: the viscosity $\gamma$, 
the ratio  $\lambda = t_s/t_o$ between the characteristic period of oscillation
$t_s = \sqrt{k_n/m_0}$ and the loading time $t_0$, the friction
coefficient $\mu$, and the ratio $\zeta=k_t/k_n$ between the tangential
$k_t$ and normal $k_n$ stiffness of the interparticle contacts.  

The viscosity factor $\gamma$ is related to the normal restitution coefficient \cite{lui}.
It was taken large enough to have a high dissipation, but not too large to 
keep the numerical stability of the method. The ratio $\lambda$ was chosen small enough in order 
to to avoid rate-dependence in the strain response, as corresponds to the 
quasi-static approximation. Technically, that is done by looking for the value of $\lambda$ such 
that a reduction of it by half makes a change of the strain response less than $5\%$.

The two parameters $\zeta$ and $\mu$ determine the constitutive response of the
system. For example, the micro-mechanical analysis of the strain response shows that the 
Young's modulus and Poisson's ratio depend on $\zeta$ \cite{Kruyt}. In the other hand, 
$\mu$ can be directly related to the friction angle of the material \cite{oger}.  
Although the study of the dependence of the constitutive response on those parameters 
is an important point, such quantities have been kept fixed in this work.

The boundary conditions yield more dimensional parameters. The initial hight
$H_0$ and width $W_0$ of the sample, and the characteristic length $\ell_0$ of the 
polygons define two geometrical parameters, which are the shape ratio $W_0/H_0$ and 
the granularity $\ell_0/H_0$ of the sample.  
 
In order to keep overlaps much smaller than the characteristic area of the polygons.
The ratio $\sigma_i/k_n$ between the stress applied on the membrane and the stiffness 
of the contacts is restricted to small values. This was implemented by fixing the 
contact  stiffness to a value closed to the experimental granular stiffness $
k_n=160MPa$. Then the stress is chosen in such way that it does not excess 
$1\%$ of this value.
 
\begin{figure}
\begin{tabular}{|c|c|c|}\hline
adimensional variable      & ratio    & default  value\\ \hline\hline
viscosity        & $\gamma $          & $0.1$\\ 
friction coefficient  & $\mu$              & $0.25$\\ 
time's ratio     & $\lambda = t_s/t_o$& $8.0\times 10^{-4} $ \\
stiffness ratio  & $\xi = k_t/k_n$    & $0.33$ \\ 
granularity      & $\ell_0/H_0$       & $0.1$ \\
shape ratio      & $W_0/H_0$          & $1.0$  \\
bending angle    & $\theta_{th}$      & $0.25\pi$\\ \hline
\end{tabular}
\end{figure}
 
\section{Stress-strain Calculation}

\subsection{Theorical background}

The macroscopic state of the system is characterized by the stress tensor and the void 
radio $e$. The area fraction of voids in the sample defines the void ratio.
Initially $e_0=0$ due to the Voronoi tessellation used.
The stress controlled test was restricted to stress states without-off diagonal
components. The diagonal component, the axial $\delta_1$ and lateral $\delta_3$ 
stress,  define the stress vector:

\begin{equation}
\tilde{\sigma}=\left[ \begin{array}{c}  p\\q \end{array} \right]
  = \frac{1}{2}\left[ \begin{array}{c} \delta_1+\delta_3 \\
                                       \delta_1-\delta_3  \end{array} \right]     
\label{stress}
\end{equation}

\noindent
where $p$ and $q$ are the pressure and the shear stress. The domain of admissible 
stresses is bounded by the failure surface. When the system reaches this 
surface it becomes unstable and fails.
 
Before failure, the constitutive behavior can be obtained performing small
changes in the stress and evaluating the resultant deformation.  An infinitesimal 
change of the stress vector $d\tilde{\sigma}$ produces an infinitesimal deformation 
of the sample, which is given by a change of height $dH$ and  width $dW$. This defines
the axial strain $d\epsilon_1 = dH/H $ and lateral strain $d\epsilon_3 = dW/W$ increments.
The volumetric strain $d\epsilon_v$ and  the shear strain  $d\epsilon_{\gamma}$
increments define the incremental strain vector:
 
\begin{equation}
d\tilde{\epsilon}=\left[ \begin{array}{c} d\epsilon_v\\d\epsilon_{\gamma} \end{array} \right]
  = \left[ \begin{array}{c} d\epsilon_1+d\epsilon_3 \\
                                       d\epsilon_1-d\epsilon_3  \end{array} \right]     
\label{strain}
\end{equation}

Each state of the sample is related to a single point in the stress space, and
the quasi-static evolution of the system is represented by the movement of this point
in the stress space.  The constitutive relation is formulated taking the incremental 
strain  as a function of the incremental stress and the stress state.

\begin{equation}
d\tilde{\epsilon}={\cal F} (d\tilde{\sigma},\tilde{\sigma})
\label{ce}
\end{equation}

If there is no rate - dependence in the constitutive equation, ${\cal F} (d\tilde{\sigma})$ 
is an homogeneous function of degree one. In this case, the application of the Euler identity 
\cite{darve} shows that Eq. (\ref{ce}) can be reduced to.

\begin{equation}
d\tilde{\epsilon}=M(\hat{\theta},\tilde{\sigma})d\tilde{\sigma}
\label{dce}
\end{equation}

\noindent
Where $\hat{\theta}$ is the unitary vector defining a specific direction in the stress
space:

\begin{equation}
\hat{\theta}= \frac{d\tilde{\sigma}}{|d\tilde{\sigma}|}
\equiv \left[ \begin{array}{c}  \cos \theta\\ \sin \theta \end{array} \right],
 ~~|d\tilde{\sigma}|=\sqrt{dp^2+dq^2}
\label{dir}
\end{equation}

The constitutive relation results from the calculation of $d\tilde{\epsilon}(\theta)$,
where each value of $\theta$ is related to a particular mode of loading. Some special 
modes are listed in the table:

\begin{figure}
\begin{tabular}{|r|c|rr|}
\hline
$ 0^o    $ & isotropic compression & $dp>0 $       & $dq=0$        \\         
$ 45^o   $ & axial loading          & $d\sigma_1>0$& $d\sigma_3=0$ \\         
$ 90^o   $ & pure shear             & $dp=0$       & $dq>0$        \\  
$ 135^o  $ & lateral loading        & $d\sigma_1=0$& $d\sigma_3>0$ \\
$ 180^o  $ & isotropic expansion    & $dp<0$       & $dq=0$        \\
$ 225^o  $ & axial stretching       & $d\sigma_1<0$& $d\sigma_3=0$ \\         
$ 270^o  $ & pure shear             & $dp=0$       & $dq<0$        \\     
$ 315^o  $ & lateral stretching     & $d\sigma_1=0$& $d\sigma_3<0$ \\ \hline  
\end{tabular}
\end{figure}

The relation (\ref{dce}) has been proposed by Darve \cite{darve} , and it contains all 
the possible  constitutive equations.  In order to interpret our particular results, it is convenient to 
make some approximations: First, if the load increments are taken small enough, the tensor 
$M(\theta)$ can be supposed to be lineal in each stress direction. Then, we assume that the
strain can be separated in an elastic (recoverable) and a plastic (unrecoverable) component:

\begin{equation}
d\tilde{\epsilon}  = d\tilde{\epsilon}^e+ d\tilde{\epsilon}^p
\label{elastoplastic}
\end{equation}

\begin{equation}
d\tilde{\epsilon}^e= D(\tilde{\sigma})d\tilde{\sigma}
\label{elastic}
\end{equation}

\begin{equation}
d\tilde{\epsilon}^p= J(\theta,\tilde{\sigma})d\tilde{\sigma}
\label{plastic}
\end{equation}

Here, $D^{-1}$ defines the stiffness tensor, and $J=M-D$ the flow rule of plasticity,
which results from the calculation of  $d\tilde{\epsilon}^e(\theta)$  and 
$d\tilde{\epsilon}^p(\theta)$. 

\begin{figure}
 \begin{center}
 \epsfig{file=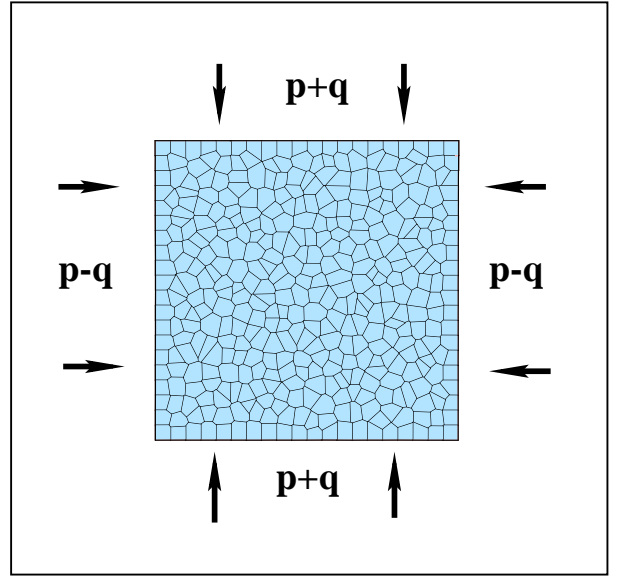,width=8.0cm,angle=0,clip=1}
 \end{center}
 \caption{Axial stress $\sigma_1=p+q$ and lateral stress $\sigma_3=p-q$ in a stress controlled test. 
They are applied on the boundary of the tessellated sample of polygons. }
 \label{voronoi}
\end{figure}

\subsection{The method}
\label{method}

\begin{figure}
 \begin{center}
 \epsfig{file=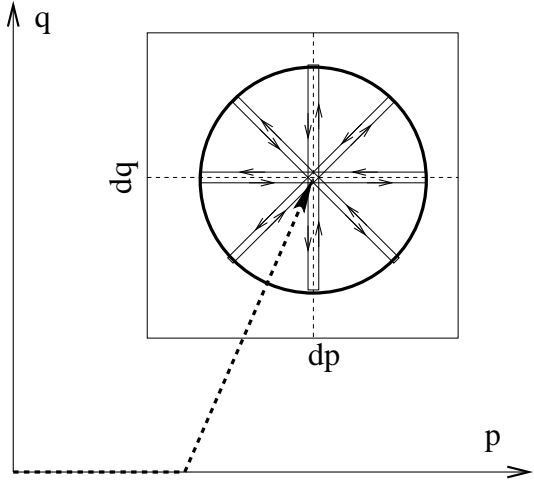,width=7.0cm,angle=0,clip=1}
 \end{center}
 \caption{Procedure to obtain the constitutive behavior: 1) The sample is driven to the 
stress state $\tilde{\sigma}$, with pressure $p$ and shear stress $q$. 2) It is loaded from
$\tilde{\sigma}$  to  $\tilde{\sigma}+d\tilde{\sigma}$  3) It is unloaded to the original stress state
$\tilde{\sigma}$.}
 \label{cr1}
\end{figure}

A numerical method has been implemented in order to find the elastic $d\tilde{\epsilon}^e$ 
and  plastic  $d\tilde{\epsilon}^p$ components of the strain as function of the stress state 
$\tilde{\sigma}$ and the stress direction $\theta$. 
Fig. \ref{cr1} shows the three steps of the procedure:

1) The sample is driven to the stress state $\tilde{\sigma}$. First, it is 
isotropically compressed until it reaches the stress value $\delta_1=\delta_3=p-q$. 
Next, it is subjected to axial loading, in order to increase the axial 
stress $\delta_1$ until $p+q$ (see Fig. \ref{voronoi}).
When the stress state $\tilde{\sigma}=[p~q]^T$ is reached, ($A^T$ being the transpose of $A$) 
the sample is allowed to relax. 

2) Loading the sample from $\tilde{\sigma}$  to  $\tilde{\sigma}+d\tilde{\sigma}$   
the strain increment $d\tilde{\epsilon}$ is obtained. This procedure is implemented
choosing different stress directions according to Eq. (\ref{dir}). Here the stress modulus  
is fixed to $|d\tilde{\sigma}|=10^{-4} p$. 

3) The sample is unloaded until the original stress state $\tilde{\sigma}$ is reached.
Then one finds a remaining strain $d\tilde{\epsilon}^p$, that corresponds to the plastic 
component of the incremental strain. Since the stress increments are taken small enough,
the unloaded stress-strain path is practically elastic. Thus, the difference  
$d\tilde{\epsilon}^e = d\tilde{\epsilon}- d\tilde{\epsilon}^p$ represents the elastic 
component of the strain.  

One could be concerned about the dependence of the strain response on the way how the stress 
state is reached.  We found that there is not remarkable dependence of the  
strain response on the stress path, whenever the stress components are quasi - static 
and  monotonically increased. Otherwise, a strong reduction in the plastic component 
of the strain is observed. In fact, when the plastic response is calculated after
the sample is unloaded, the plasticity results smaller than that one calculated after a 
monotonic load.   Furthermore, there is no plastic component in the strain response
when elastic waves are previously generated in the sample. Those memory effects suggest 
that the plastic component of the strain depend on the history of the deformation, and is 
kept unchanged only if the sample is subjected to quasi-static and monotonic load.

\begin{figure}
 \begin{center}
 \epsfig{file=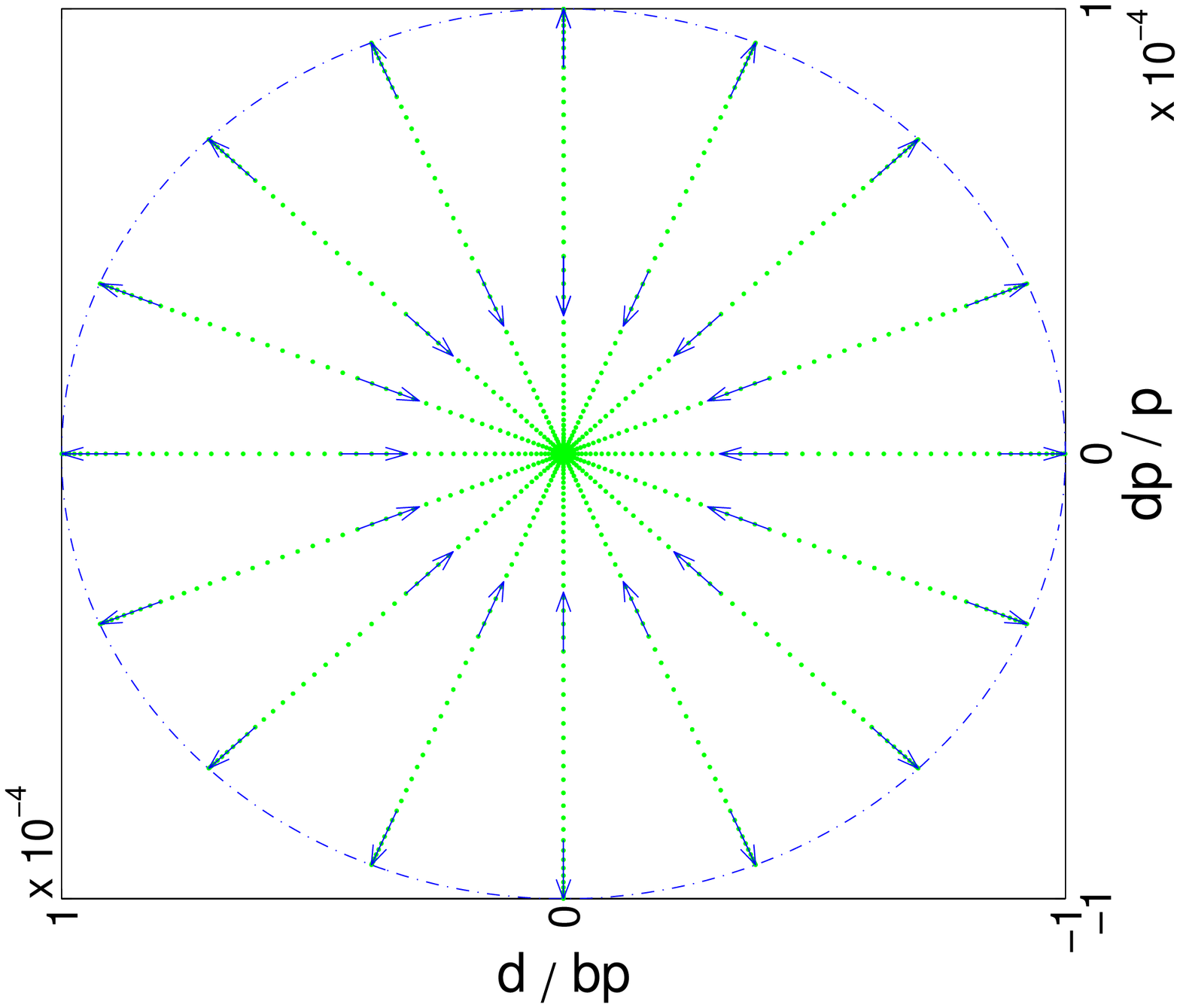,width=6.5cm,angle=-90,clip=1}
 \epsfig{file=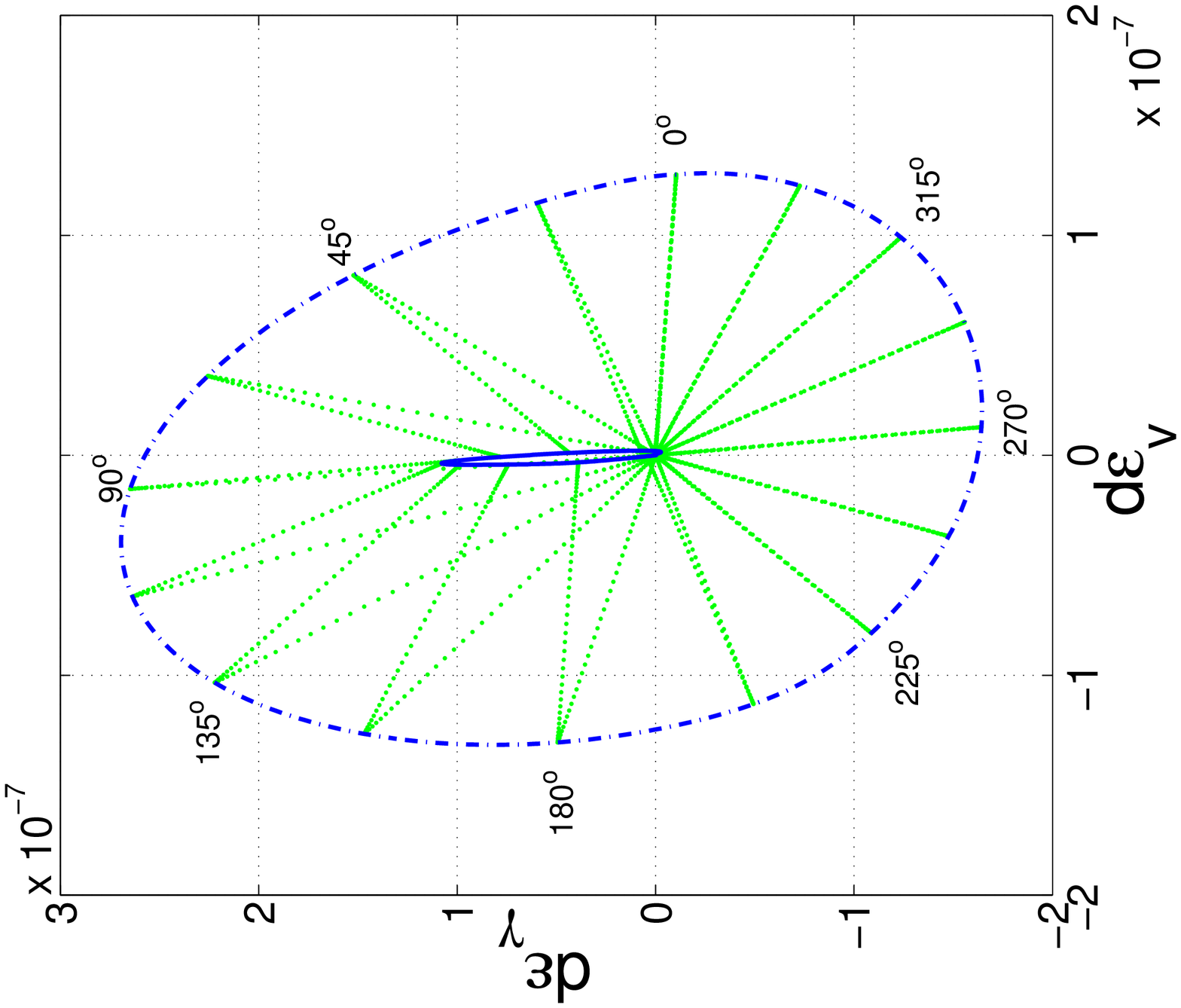,width=6.3cm,angle=-90,clip=1}
 \end{center}
\caption{Stress - strain relation resulting from the load - unload test. Dotted lines
represent the paths in the stress and strain spaces. The dash-dot line gives the strain 
envelope response and the solid line is the plastic envelope response.}
\label{de1}
\end{figure}

Fig. \ref{de1} shows the load-unload paths and the corresponding strain response. They were
taken from a stress-state with $q = 0.5 p$.  The end of the load
paths in the stress space map into a strain envelope response $d\tilde{\epsilon}(\theta)$
in the strain space. Likewise, the end of the unload paths map into a plastic envelope 
response $d\tilde{\epsilon}^p(\theta)$. The yield direction $\phi$ can be found from 
this response, as the direction in the stress space where the plastic response is maximal. 
The flow rule can be obtained taking the direction $\psi$ of the maximal plastic response 
in the strain  space. These angles do not agree, that reveals the necessity to analyze this
behavior in the framework of the non-associated theory of plasticity (see Sect. \ref{flow}).

\section{Constitutive Relation}
\label{cr}

\begin{figure}
 \begin{center}
 \epsfig{file=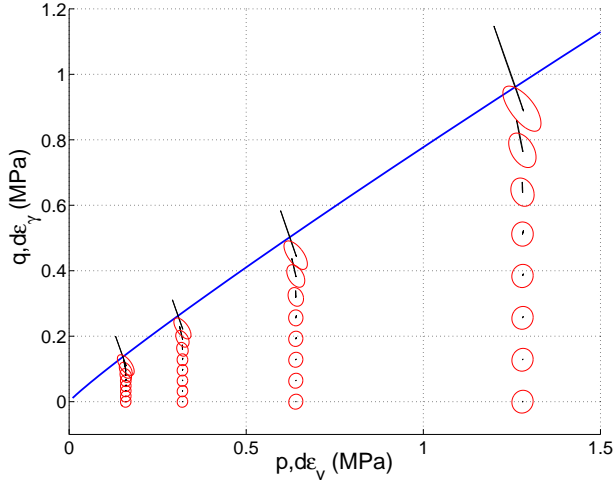,width=6.5cm,angle=-90,clip=1}
 \end{center}
 \caption{Elastic response $d\tilde{\epsilon}^e$ and plastic response $d\tilde{\epsilon}^p$  
resulting from the application of different loading modes with $|d\tilde{\sigma}|=10^{-4} p$. 
The solid line represents the failure surface.}
 \label{elpl}
\end{figure}

Fig.\ref{elpl} summarizes the global elasto-plastic behavior.  The elastic response, 
calculated  from Eq. (\ref{elastoplastic}), has a centered ellipse as envelope response. 
This can be related to the micro-contact structure using a local linear relation in each point 
of the stress space (see Sect.\ref{stiffness}). The solid line represents the failure surface, 
which separates  the stable states of the unstable ones (see Sect. \ref{failure}). The 
plastic envelope response is almost on a straight line. The modulus and the orientation of 
this envelope depend on the stress state through a certain number of material parameters, which
are given in Sect. \ref{flow}.  All the quantities obtained in this section have been calculated 
from the average over five different samples of $ 10 \times 10 $  particles each one.   

\subsection{Failure surface}
\label{failure}

\begin{figure}
 \begin{center}
 \epsfig{file=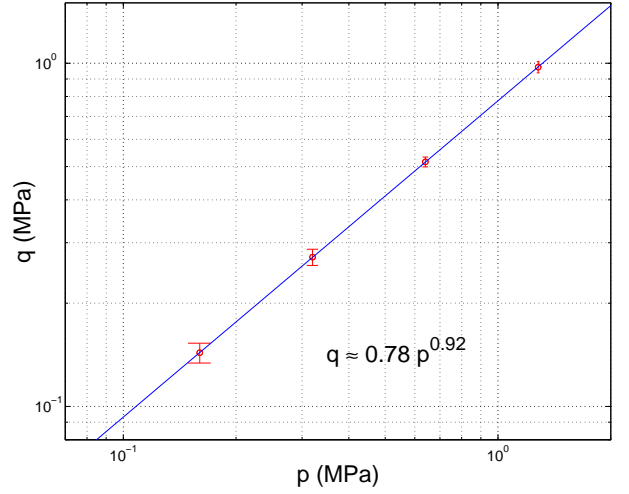,width=6.5cm,angle=-90,clip=1}
 \end{center}
 \caption{Failure surface. The continuous line represents the power law fit. }
 \label{fail}
\end{figure}

The failure line was calculated looking for the values of stress for which the system becomes unstable:
for each pressure $p$, there is a critical shear stress $q_c(p)$,
below which the sample reaches a stable state with an exponential decay of its 
kinetic energy. For shear stress values above the critical one, the sample develops an instability 
and fails. Fig. \ref{fail} shows the interface between these two stress states, that can be accurately 
fitted by the power law:

\begin{equation}
\frac{q_c}{p_0}= \mu^*\left(\frac{p}{p_0}\right)^{\beta} 
\label{fline}
\end{equation}

\noindent
Here $p_0=1.0MPa$ is the reference pressure, and $\mu^*=0.78\pm0.03$ is the Mohr-Coulomb
friction coefficient \cite{vermeer}. The power law dependence on the pressure, with exponent 
$\beta = 0.92 \pm 0.02 $ implies a significant deviation from the Mohr-Coulomb theory.
Moreover, the empirical criteria of failure for most rocks \cite{cook} shows 
a power law dependence of the form of Eq. (\ref{fline}). It seems that additional features 
beyond the Mohr-Coulomb analysis are taking place when the sample fails, that will be discussed in 
sect. \ref{flow}. 

\subsection{stiffness}
\label{stiffness}

Hooke's law of elasticity states that the stiffness tensor of isotropic materials can be
written in terms of two material parameters: the Young's modulus $E$ and the Poisson's ratio 
$\nu$.  However, the isotropy is not fulfilled when the stress state is far from the hydrostatic 
axis. Indeed, numerical simulations \cite{thornton,mark} and photo-elastic experiments \cite{oda} on 
granular materials show that the loading induces a significant departure from isotropy in 
the contact  network.

The anisotropy of the granular sample can be characterized by the distribution of the micro-contact
normal vectors ${\hat{n}^c_i}$ (see Fig. \ref{overlap}). Our numerical simulations show that 
the structural changes of micro-contacts are principally due to the opening of contacts whose
normal vectors are nearly aligned around the direction perpendicular to the load.  
Let us call $N(\varphi)\Delta\varphi$  the  number of contacts per particle, oriented between the 
angles $\varphi$ and  $\varphi + \Delta\varphi$, measured with respect to the direction in which 
the sample is loaded. The lowest order of anisotropy can be  described by:

\begin{equation}
N(\varphi) = \frac{1}{2\pi}\big[N + (N_0-N)\cos(2\varphi)\big]
\label{pdc}
\end{equation}

\begin{figure}
 \begin{center}
 \epsfig{file=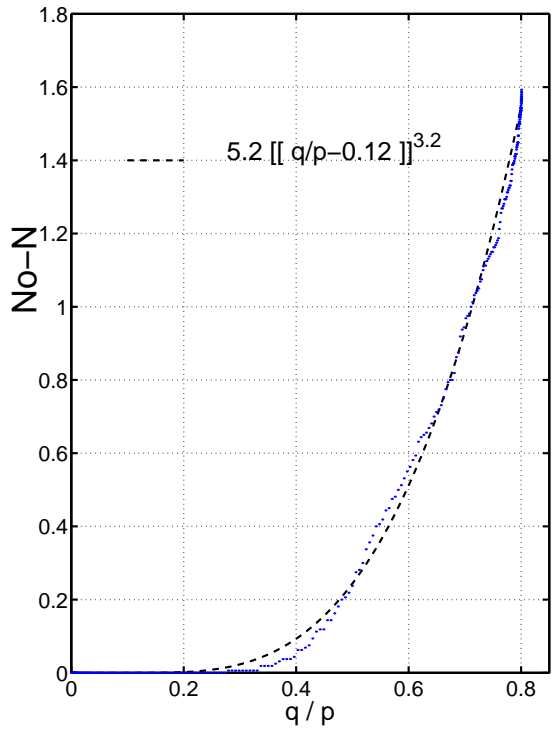,width=7.3cm,angle=0,clip=1}
 \end{center}
 \caption{Reduction of the mean coordination number of contacts (dotted line). The data have been
fitted to a truncated power law (dashed line). See Eq. (\ref{trunc}). }
 \label{damage}
\end{figure}

Here $N$ is the average coordination number of the polygons, whose initial value $N_0 = 6.0 $
reduces as the load is increased.  Fig. \ref{damage} shows this reduction. A critical line is found 
around  $q=0.12p$, below which there are no structural changes in the contact network.  Above
this limit an induced anisotropy arises due to opened contacts whose amount follows a power law 
dependence.

In order to describe the effect of the anisotropy in the elastic response we proceed as 
follows:  first, an additional parameter $\alpha$ is included in Hooke's law 

\begin{equation}
\left[ \begin{array}{c}
d\epsilon^e_1 \\
d\epsilon^e_3
\end{array} \right]
=\frac{1}{E}\left[ \begin{array}{cc}
1-\alpha    & -\nu \\
-\nu &  1+\alpha\end{array} \right]
\left[ \begin{array}{c}
d\sigma_1 \\
d\sigma_3
\end{array}\right]
\label{hooke}
\end{equation} 

\noindent
Then, these three parameters are supposed to be depended on the internal damage parameter $d$:

\begin{equation}
d=\frac{N_0-N}{N_0}
\end{equation}

The tensor $D$ defined in Eq. (\ref{elastic}) is calculated from  Eq. (\ref{hooke}) using the definition 
of the stress and strain vectors given in Eqs. (\ref{stress}) and (\ref{strain}). One obtains:

\begin{equation}
D=\frac{2}{E}\left[ \begin{array}{cc}
                  1-\nu&   -\alpha  \\
                  -\alpha  &   1+\nu  
\end{array} \right]
\label{D}
\end{equation}

\noindent
The diagonal components of this tensor are respectively the inverse of the bulk modulus and of the 
shear modulus. The non-diagonal component results from the anisotropy of the sample, and it
couples the compression mode with the shearing deformation. These three variables are calculated 
from the elastic response $d\tilde{\epsilon}^e(\theta)$ by the introduction of the following
function:

\begin{equation}
R(\theta)=\frac{d\tilde{\sigma}^T d\tilde{\epsilon}^e }{|d\tilde{\sigma}|^2}
\label{R}
\end{equation}

\noindent

by substitution of Eqs. (\ref{elastic}) and (\ref{dir}) into Eq. (\ref{R}), one sees that $R$ 
is the quadratic form of $D$:

\begin{equation}
R(\theta)= \hat{\theta}^TD\hat{\theta} 
= \frac{2}{E}\big[1-\nu\cos(2\theta)-\alpha\sin(2\theta) \big]
\end{equation} 

\noindent
Using this equation, the components of $D$ can be evaluated as the Fourier coefficients of $R$:

\begin{eqnarray}
\frac{1}{E}& = & \frac{1}{4\pi}\int_0^{2\pi} R(\theta) d\theta\\
        \nu& = &-\frac{E}{2\pi}\int_0^{2\pi} R(\theta)\cos(2\theta) d\theta\\
     \alpha& = &-\frac{E}{2\pi}\int_0^{2\pi} R(\theta)\sin(2\theta) d\theta
\end{eqnarray}  

\begin{figure}
 \begin{center}
 \epsfig{file=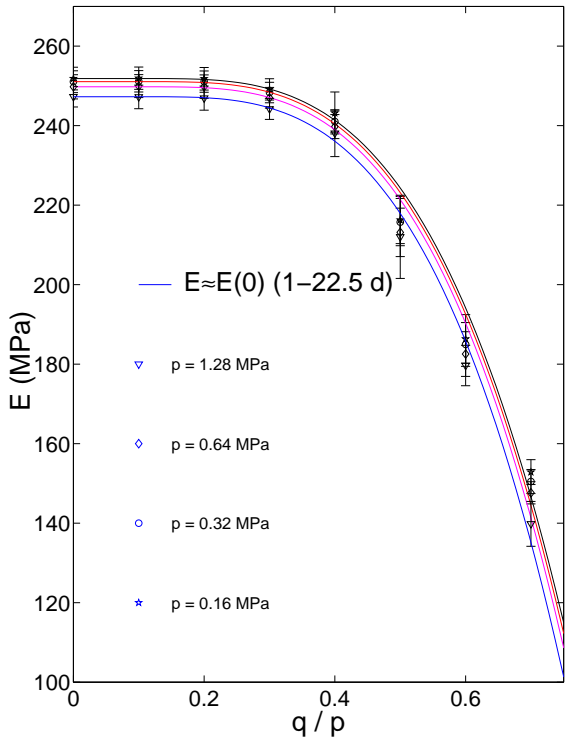,width=7.5cm,angle=0,clip=1}
 \end{center}
 \caption{Young's modulus. The solid line is the linear approximation of $E(d)$. See Eq. (\ref{tay}).}
 \label{young}
\end{figure}

\begin{figure}
 \begin{center}
 \epsfig{file=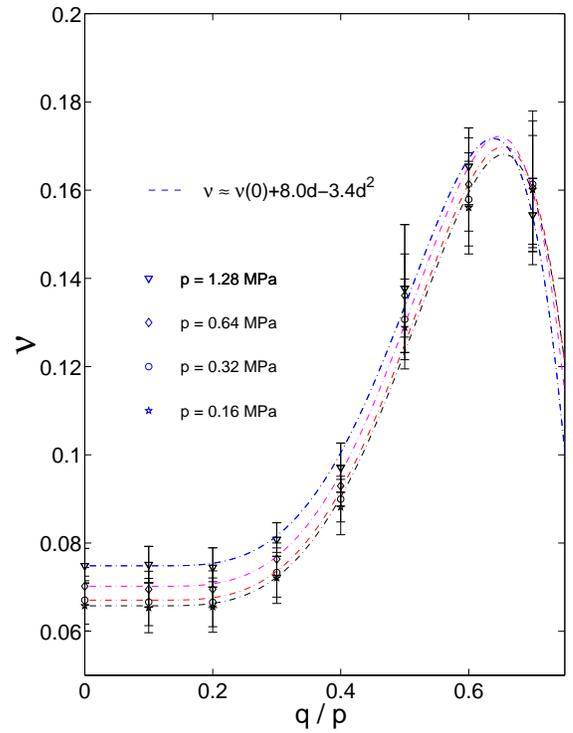,width=7.5cm,angle=0,clip=1}
 \end{center}
 \caption{Poisson's ratio. The dashed line is the quadratic approximation of $\nu(d)$. See Eq. (\ref{tay}).}
 \label{poisson}
\end{figure}

\begin{figure}
 \begin{center}
 \epsfig{file=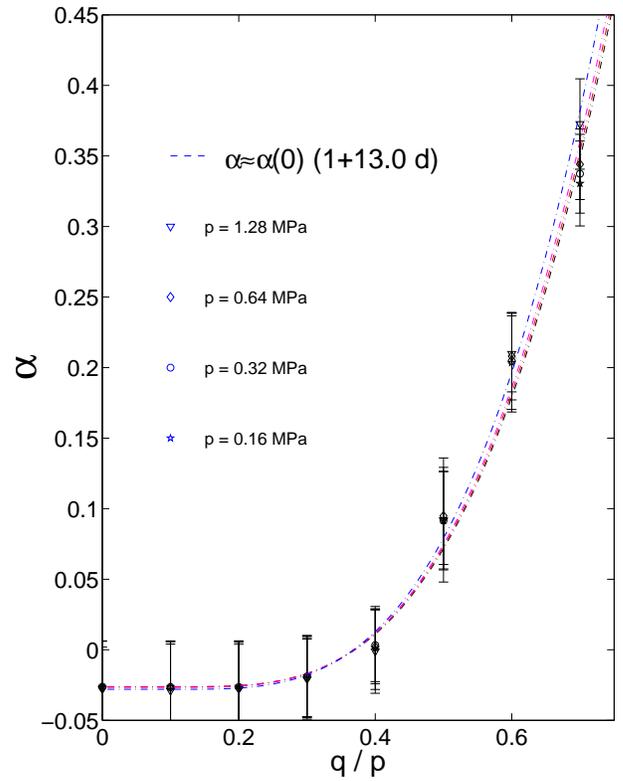,width=8.0cm,angle=0,clip=1}
 \end{center}
 \caption{Anisotropy parameter. The dashed line is the linear approximation of $\alpha(d)$. See Eq. (\ref{tay}).}
 \label{aniso}
\end{figure}

Figs. \ref{young}, \ref{poisson} and \ref{aniso} show the results of the calculation of 
the Young's modulus $E$, the  Poisson's ratio $\nu$ and the anisotropy factor $\alpha$.
Below the limit of isotropy, Hooke's law can be applied: $E\approx E_0$, $\nu \approx \nu_0$ 
and $\alpha \approx 0$. On the other hand, above the limit of isotropy a reduction of the 
Young's modulus is found, along with an increase of the Poisson's ratio and the 
anisotropy factor. The functional dependence of those parameters with the internal damage 
parameter $d$ are evaluated developing their Taylor's series around $d=0$:

\begin{eqnarray}
E(d)     &=& E(0)+E'(0)d+                O\left(d^2\right) \nonumber \\
\alpha(d)&=& \alpha (0)+\alpha'(0)d+     O\left(d^2\right) \\
\label{tay}
\nu(d)   &=& \nu(0)+\nu'(0)d+\nu''(0)d^2+O\left(d^3\right) \nonumber
\end{eqnarray}

The coefficients of this expansion are calculated from the best fitting of those expansions.
Figs. \ref{young} and \ref{aniso} show that the linear approximation is good enough
to reproduce the Young's modulus and the anisotropy factor. The fit of the Poisson's ratio, 
however, requires the inclusion of a quadratic approximation,  implying that it has a non-linear
dependence on the damage parameter (Fig. \ref{poisson}).

\subsection{Plastic Flow}
\label{flow}

The formulation of the non-associated theory of plasticity requires the evaluation of
three material functions: the yield direction $\phi$, the flow direction $\psi$, and the 
plastic modulus $h$. These quantities can be calculated from the plastic response 
$d\tilde{\epsilon}^p(\theta)$, as follows:

The yield direction is given by the incremental stress direction $\phi$ with maximal
plastic response:

\begin{equation}
|d\tilde{\epsilon}^p(\phi)|=\max_{\theta}{|d\tilde{\epsilon}^p(\theta)|}
\label{yd}
\end{equation} 

\noindent
The flow direction is defined from the orientation of the plastic response at its maximum value:

\begin{equation}
\psi = \arctan(\frac{d\epsilon^p_{\gamma}}{d\epsilon^p_v})\left|_{\theta=\phi}\right.     
\label{fd}
\end{equation}

\noindent
The plastic modulus is obtained form the modulus of the maximal plastic response:

\begin{equation}
\frac{1}{h} = \frac{|d\tilde{\epsilon}^p(\phi)|}{|d\tilde{\sigma}|}
\label{hd}
\end{equation} 

\noindent

Reciprocally, the plastic response can be expressed in terms of these quantities. 
Let us define the unitary vectors $\hat{\psi}$ and $\hat{\psi}^{\perp}$. The first one 
is oriented in the direction of $\psi$ and the second one is the $90^o$ rotation of $\hat{\psi}$. 
The plastic strain is written in this basis as:

\begin{eqnarray}
d\tilde{\epsilon}^p(\theta) &=& [(d\tilde{\epsilon}^p)^T\hat{\psi}]\hat{\psi}
+[(d\tilde{\epsilon}^p)^T\hat{\psi}^{\perp}]\hat{\psi}^{\perp} \nonumber\\ 
&\equiv&  \frac{1}{h}\left[ f(\theta)\hat{\psi} + g(\theta)\hat{\psi}^{\perp} \right]
\label{pl1}
\end{eqnarray}


\newpage

\begin{figure}
 \begin{center}
 \epsfig{file=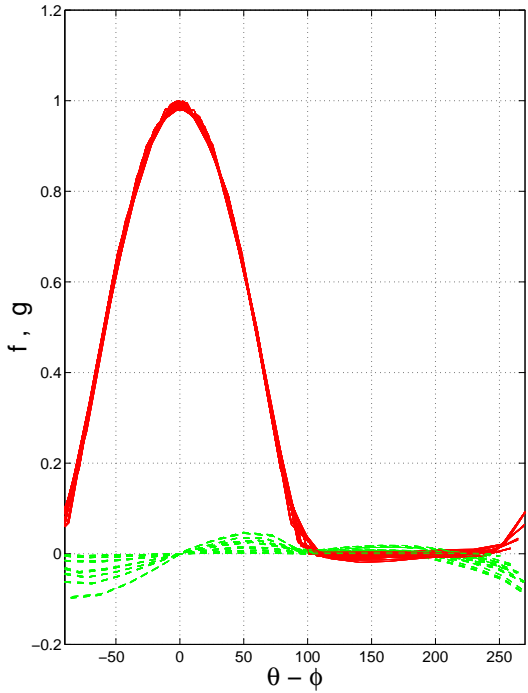,width=7.0cm,angle=0,clip=1}
 \end{center}
 \caption{Plastic profiles $f(\theta)$ (solid line) and $g(\theta)$ (dashed line). The results
for different stress values have been superposed. }
 \label{profile}
\end{figure}

The plastic profiles $f(\theta)$ and $g(\theta)$ are shown in Fig. \ref{profile}.
The first one is approximately the same for all the stress states, and can be well fitted
to a cosine function, centered on the yield direction $\phi$ and truncated to zero for the negative
values. The last profile depends on the stress value, and is difficult to evaluate, because 
it is of the same order as the statistical fluctuations. However, the 
contribution of $g$ to the total strain response is negligible. 
In order to simplify the description of the plastic response, 
the following approximation is made: 

\begin{equation}
g(\theta)\ll f(\theta)\approx [[cos(\theta-\phi)]]=[[\hat{\phi}^T\hat{\theta}]]
\label{pl2}
\end{equation}

\noindent
where $[[\cdot]]$ defines the function

\begin{equation}
 [[x]]= \left\{ \begin{array}{r@{\quad:\quad}l} 
                  0 & x \le 0 \\ x & x > 0
                  \end{array} \right. 
\label{trunc}
\end{equation}

Now, The flow rule results from the substitution of Eqs.(\ref{pl1}) and (\ref{pl2}) into Eq.
(\ref{plastic}):

\begin{equation}
J(\theta)d\tilde{\sigma}=\frac{[[\hat{\phi}^Td\tilde{\sigma} ]]}{h}\hat{\psi}
\label{flowrule}
\end{equation}

\begin{figure}
 \begin{center}
 \epsfig{file=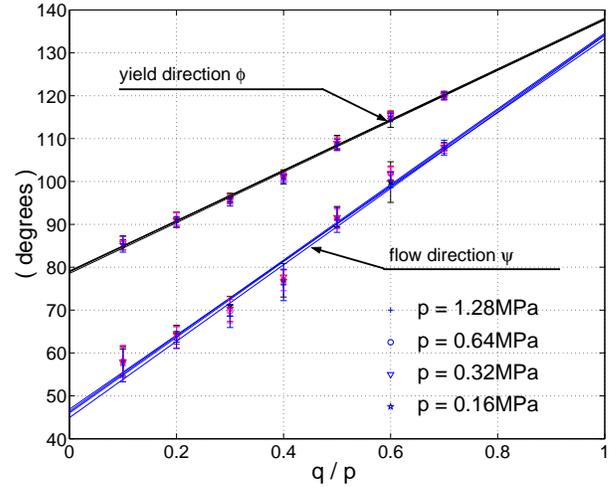,width=6.5cm,angle=-90,clip=1}
 \end{center}
 \caption{The flow direction and the yield direction of the plastic response. Solid lines
represent a linear fit. }
 \label{phipsi}
\end{figure}

The yield direction and the flow direction have been calculated for different stress states.
The results are shown in  Fig. \ref{phipsi}.  Both angles are quite different, which 
is a clear deviation from Drucker's normality postulate \cite{drucker}.
Indeed, many experimental results on soil 
deformation \cite{tatsouka} have confirmed that these angles are completely 
different. Thus Drucker's postulate is not fulfilled in the deformation
of granular materials, and the main reason for that is the rearrangement of contacts 
on small deformations which are not taken into account in this theory. 
On the other hand, all the sliding, opening, and  other micro-mechanical 
rearrangements can be well handled in the discrete element formulation, 
which results more adequate to describe the soil deformation.

The material constants are  evaluated from the dependence of the plastic quantities 
on the stress:  the yield direction and the flow direction can be roughly approximated by
straight lines: 

\begin{eqnarray}
\phi &=& \phi_0 +\phi'_0\frac{q}{p}\nonumber\\
\psi &=& \psi_0 +\psi'_0\frac{q}{p}
\label{lfr}
\end{eqnarray}

\begin{figure}
 \begin{center}
 \epsfig{file=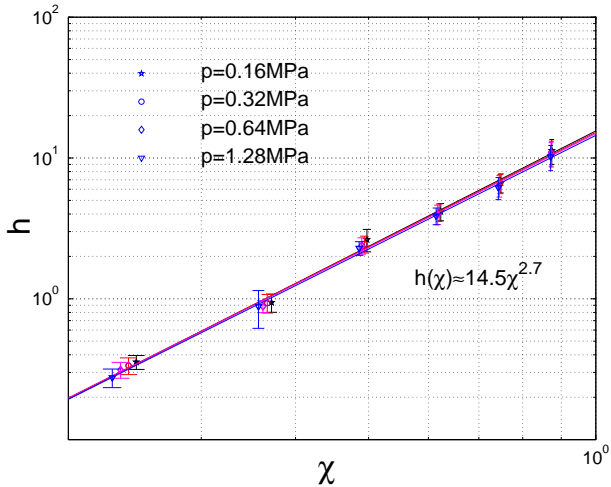,width=6.5cm,angle=-90,clip=1}
 \end{center}
 \caption{Plastic modulus. The solid line is a power law fit with respect to the variable
 $\chi= 1-(p/p_0)^{\vartheta}q/(\mu^* p_0)$. }
 \label{hard}
\end{figure}

The four material parameters $\phi_0 =46^o \pm 0.75^o$, $\phi'_0 = 88.3^o \pm 0.6^o$, 
$\psi_0 = 78.9^o \pm 0.2^o$ and $\psi'_0 = 59.1^o \pm 0.4^o$ are obtained
from the linear fit of the data. On the other hand, Fig. \ref{hard} 
shows that the plastic modulus depends on the stress through a power law relation: 

\begin{equation}
h= h_0 \left[ 1 - \frac{q}{\mu^* p_0}(\frac{p_0}{p})^{\vartheta}\right]^{\eta}
\label{hardening} 
\end{equation}

\noindent
There are four additional material parameters: The plastic modulus $h_0 = 14.5 \pm 0.05$ 
at $q=0$, the Mohr-Coulomb friction coefficient $\mu^*$ (see Eq. (\ref{fline})),
and the exponents $\eta = 2.7 \pm 0.04 $ and  $\vartheta =  0.981 \pm 0.002$ . 

The plastic limit surface is given by the stress states where the plastic deformation
becomes infinite. According to the flow rule ( Eq. (\ref{flowrule})), it is found
looking for the stress values where Eq. (\ref{hardening}) vanishes:

\begin{equation}
\frac{q_p}{p_0}= \mu^*\left(\frac{p}{p_0}\right)^{\vartheta} 
\label{pline}
\end{equation}

It is important to point out that the failure surface --given in  Eq. (\ref{fline})-- 
does not correspond to the plastic limit surface. Actually, This matter has 
already been discussed in the framework of the Hill's condition of instability 
\cite{darve2} the bifurcation analysis \cite{vardoulakis}, which predict that the instability 
should be reached strictly inside the plastic limit surface.

\section{CONCLUDING REMARKS}

The elasto-plastic response of a Voronoi tessellated sample of polygons has been calculated 
in the case of monotonic and quasi-static loading. It can be written in a simple form as:

\begin{equation}
d\tilde{\epsilon}=D(d)d\tilde{\sigma}+\frac{[[\hat{\phi}^Td\tilde{\sigma} ]]}{h}\hat{\psi}
\label{monotonic}
\end{equation}

The plastic response reflects the non-associated features of  realistic soils. Here the yield
direction and flow direction are linearly related to  the ratio $q/p$, and the plastic modulus 
obeys a power law relation with a weak pressure dependence.

The classical parameters of elasticity -- the Young's modulus and the Poisson's ratio -- are not 
material constants, because they depend on the internal damage parameter. Therefore, 
Eq. (\ref{monotonic}) is not complete, and it is necessary to include a constitutive equation
that relates the internal damage to the external load. By focusing on the details of the 
dynamic of the micro-contacts, significant progress may be made in the macroscopic description
of the deformation.

\begin{figure}
 \begin{center}
 \epsfig{file=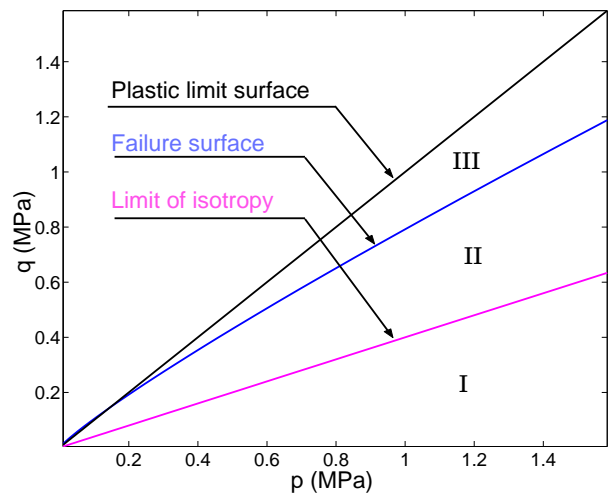,width=7.0cm,angle=-90,clip=1}
 \end{center}
 \caption{Elasto-plastic regimes: isotropic(I), anisotropic (II) and unstable (III). }
 \label{phase}
\end{figure}

The elasto-plastic response leads to the identification of three different regimes which are shown
in  Fig. \ref{phase}. Zone I corresponds to the isotropic regime, characterized by small plastic 
deformations and a linear elastic regime.  In the zone II open contacts are detected, which must 
be taken into account in the calculation of the non linear elasticity. Zone III corresponds to 
unstable states so that the stress-strain relation can not be calculated here.  The extrapolation 
of the strain response in this region shows that the plastic strain must have a finite value just before 
the instability is reached.

The above observation leads to the open question of the nature of the failure \cite{darve2}. 
Numerical simulations on strain controlled tests show that strain localization is the most 
typical mode of failure. The fact that it appears before the sample reaches the plastic limit
surface suggests that the appearance of the instability is not completely  determined by the 
macroscopic state.  

The role of the microstructure on the strain localization  has been intensely studied in the last
years\cite{vardoulakis,desrues}. Future work is the creation of samples with different granular 
textures --for example, changing the void ratio distributions and the polydispersity
of the grains--. Then we can deal with the question of how a change in the microstructure 
affects the elasto-plastic response and the strain localization.

\section*{Acknowledgments}

We thank F. Darve, P. Vermeer, F. Kun, J. \AA str\o m and S. Luding for helpful discussions
and acknowledge the support of the {\it Deutsche Forschungsgemeinschaft\/} within the research group 
{\it Modellierung koh\"asiver Reibungsmaterialen\/}.

\end{multicols}
\end{document}